\documentclass[a4paper,11pt]{article}
\pdfoutput=1
\usepackage{jheppub}

\usepackage{mathrsfs,amsmath,amsthm,amssymb,braket,verbatim,adjustbox,multirow,enumerate,textcomp,gensymb,subfigure,graphicx,diagbox,pifont,tabulary,booktabs,stackengine,epstopdf,dcolumn,makecell,listings,bm,dsfont,mathtools,autobreak,slashed,float,array,colordvi,xfrac}
\usepackage[usenames,dvipsnames]{color}
\usepackage{lineno}
\modulolinenumbers[5]
\usepackage{hyperref}
\hypersetup{bookmarksopen,bookmarksnumbered,citecolor=NavyBlue,linkcolor=BlueViolet,urlcolor=[rgb]{0,0.35,0.5}}

\definecolor{eprintLinks}{rgb}{0,0.35,0.5}
\definecolor{journalLinks}{rgb}{0,0.35,0.5}
\newcommand{\MYhref}[3][blueLinks]{\href{#2}{\color{#1}{#3}}}

\def\lp{l_{\textrm{P}}}
\def\Gn{G_{\textrm{N}}}
\def\Mp{M_{\textrm{P}}}
\def\Elv{E_{\textrm{LV}}}
\def\I{\textrm{i}}
\def\bfp{{\bf p}}
\def\O{{\cal O}}
\def\bfu{{\bf u}}
\def\un{\lvert\bfu\rvert}
\def\II{{\hbox{{\rm I}\kern-.2em\hbox{\rm I}}}}
\def\e{\textrm{e}}
\def\dla{\langle\!\langle}
\def\dra{\rangle\!\rangle}
\def\pn{\lvert\bfp\rvert}
\def\d{\textrm{d}}
\def\D{{\cal D}}
\def\IC{{\hbox{{\rm I}\kern-.5em\hbox{\rm C}}}}
\def\texthollowstar{\text{\ding{73}}}

\makeatletter
\let\latex@fnsymbol\@fnsymbol
\renewcommand\@fnsymbol[1]{\ifcase#1\or\texthollowstar\or\textdagger\else\@ctrerr\fi}
\newcommand{\restorefnsymbol}{\let\@fnsymbol\latex@fnsymbol}
\makeatother

\renewcommand{\thefootnote}{\fnsymbol{footnote}}

\title{\boldmath Effects on neutrino propagation in space-time foam of D-branes revisited\footnote{Published as JHEP 05 (2024) 266, \url{https://doi.org/10.1007/JHEP05(2024)266}}}

\author[a]{Chengyi Li}\emailAdd{lichengyi@pku.edu.cn}
\affiliation[a]{School of Physics, Peking University, Beijing 100871, China}
\author[a,b,c]{and Bo-Qiang Ma\footnote{Author to whom any correspondence should be addressed.}}\emailAdd{mabq@pku.edu.cn}
\affiliation[b]{Center for High Energy Physics, Peking University, Beijing 100871, China}
\affiliation[c]{Collaborative Innovation Center of Quantum Matter, Beijing, China}

\abstract{Neutrinos from the cosmos have proven to be ideal for probing the nature of space-time. Previous studies on high-energy events of IceCube suggested that some of these events might be gamma-ray burst neutrinos, with their speeds varying linearly with their energy, implying also the coexistence of subluminal and superluminal propagation. However, a recent reanalysis of the data, incorporating revised directional information, reveals stronger signals that neutrinos are actually being slowed down compared to previous suggestion of neutrino speed variation. Thus, it is worth discussing its implications for the brane/string inspired framework of space-time foam, which has been used to explain previous observations. We revisit effects on neutrino propagation from specific foam models within the framework, indicating that the implied violation of Lorentz invariance could necessarily cause the neutrino to decelerate. We therefore argue that this sort of model is in agreement with the updated phenomenological indication just mentioned. An extended analysis of the revised IceCube data will further test these observations and stringy quantum gravity.}

\keywords{D-Branes, Non-Standard Neutrino Properties, String Models, Violation of Lorentz and/or CPT Symmetry}

\arxivnumber{2401.05867}

\begin{document}
\maketitle
\flushbottom

\renewcommand{\thefootnote}{\arabic{footnote}}

\noindent
The quantization of gravitational field would endow space-time itself a `foamy' structure at scales of order the Planck length $\lp=\sqrt{\hslash \Gn/c^{3}}\simeq 1.6\times 10^{-33}$~cm~(corresponding to the scale $\Mp=1/\lp\sim 10^{28}$~eV in units of $c=\hslash=1$ used here). This is expected to modify the propagation of ultrarelativistic particles such as neutrinos, such that they may experience a nontrivial index of refraction \textit{in vacuo}. In the often considered scenario of a linear violation of Lorentz invariance~(LV) motivated by the underlying idea of quantum gravity~(QG)~\cite{He:2022gyk}, this would result in a modification increasing linearly with the energy $E$ of the particle to the velocity of propagation, $v=1-sE/\Elv$, where $\Elv$, of order $\Mp$ from one's intuition, stands for the first-order LV scale to be measured experimentally, and, $s=+1~(-1)$, the LV sign factor, indicates the `subluminal'~(`superluminal') effect from quantized space-time.  Neutrinos from cosmologically remote objects like gamma-ray bursts~(GRBs) are suggested to be ideal for probing such candidate QG effects~\cite{Jacob:2006gn,Amelino-Camelia:2009imt,Amelino-Camelia:2015nqa}. Albeit events with energies of several hundreds of TeV~(and those even exceeding PeV energies) have been registered~\cite{IceCube:2013cdw,IceCube:2016uab} by the IceCube telescopes~\cite{Gaisser:2014foa}) one is not sure about their exact sources still.

Recent studies~\cite{Amelino-Camelia:2015nqa,Amelino-Camelia:2016fuh,Amelino-Camelia:2016ohi,Huang:2018ham,Huang:2019etr,Huang:2022xto} using IceCube data of near-TeV to PeV energies compare observation times and directions of GRBs with those of neutrinos, finding intriguing preliminary signatures that there could be a variation in the speed of GRB neutrinos. Comparable statistical significance for events delayed and those that advanced by LV is inferred,\footnote{We recall that 5 of 9~(candidate) GRB neutrinos appear to have travel times retarded~(i.e., neutrinos that could have been slowed down) by QG, whereas the other 4 are `advance' events, if confined to IceCube 60--500~TeV events~\cite{Amelino-Camelia:2016fuh,Amelino-Camelia:2016ohi}; likewise, for the four PeV neutrinos~(that include 1 `track' event and 3 `shower' events), 2 are delayed~(so subluminal) and the rest 2, superluminal, see~\cite{Huang:2018ham}, where the interested reader will find these described in detail.} pointing likely to an opposite effect on the propagation between neutrinos and antineutrinos~\cite{Huang:2018ham,Huang:2019etr,Huang:2022xto,Zhang:2018otj}~(a hypothesis that might be relevant because we have no helicity information for IceCube events). It is suggested~\cite{Li:2023wlo} that the space-time foam model with recoiling~(D-brane) cosmic defects~\cite{Ellis:1999uh} may induce a breakdown of CPT invariance for the neutrino/antineutrino sector and can accommodate~\cite{Li:2022sgs} then for such an asymmetric impact of LV on the neutrino propagation. A latest reanalysis~\cite{Amelino-Camelia:2022pja} indicates, however, that these previous results might be strongly affected by the significant revision of the directional information for the neutrino events by the collaboration~\cite{IceCube:2020wum} and shows that the prior indication for a QG effect speeding up neutrinos~(or antineutrinos) seems not to be favored anymore~(given whopping probability of finding accidentally that feature from data), whereas for neutrinos subject to subluminal LV, the result becomes even stronger than those previously reported~(using uncorrected neutrino directions) in Refs.~\cite{Amelino-Camelia:2016fuh,Amelino-Camelia:2016ohi,Huang:2018ham,Huang:2019etr,Huang:2022xto}. Indeed, as has been emphasized, therein, many times, one is still far from confirming that any of the suggested `GRB-neutrinos' were a neutrino from GRB: the association of neutrinos with GRB can of course occur purely accidentally; but if this connection becomes statistically significant enough~\cite{Amelino-Camelia:2022pja}, it would evidently be meaningful, as is likely what happened to \textit{delayed} neutrinos.

On the other hand, in addition to Refs.~\cite{Amelino-Camelia:2016fuh,Amelino-Camelia:2016ohi,Huang:2018ham,Amelino-Camelia:2022pja} in focusing on the `shower' events of IceCube, it has been found that 12 `track' events of lower energies around 1~TeV~\cite{Huang:2019etr} and another 3 or more multi-TeV tracklike neutrino signals~(above 30~TeV) are probably GRB neutrinos with retarded travel times~\cite{Huang:2022xto}, pointing also to a more significant signature for subluminal propagation of neutrinos compared to that previously found from events above 60~TeV~\cite{Amelino-Camelia:2016fuh,Amelino-Camelia:2016ohi,Huang:2018ham}. While the newly corrected directions~\cite{IceCube:2020wum} could not have been considered there~\cite{Huang:2019etr,Huang:2022xto} at the time, the findings in some sense are compatible with the new result of Ref.~\cite{Amelino-Camelia:2022pja}, suggesting a \textit{neutrino speed variation} of $v(E)=1-E/\Elv$, where $\Elv\coloneqq\Mp/\eta\sim 6\times 10^{17}$~GeV~(that is $\eta\simeq 21.7$), a scale roughly the same as that determined before by~\cite{Amelino-Camelia:2016fuh,Amelino-Camelia:2016ohi,Huang:2018ham,Huang:2019etr,Huang:2022xto}.\footnote{Such an estimate of the scale is compatible with the constraints established by the neutrino pulses of supernovae~\cite{Ellis:2008fc} as well as~(see, e.g.,~\cite{Ellis:2018ogq}) the multimessenger identification of blazar TXS~0506+056 coincident with a $\sim 290$~TeV tracklike neutrino~\cite{IceCube:2018dnn}.}

As the analysis~\cite{Amelino-Camelia:2022pja}, based on revised data, weakens superluminal signature, i.e., correlations~\cite{Amelino-Camelia:2016fuh,Amelino-Camelia:2016ohi,Huang:2018ham,Huang:2019etr,Huang:2022xto} of relevant events with GRBs probably are just coincident, it is worth reconsidering whether the finding can still be described in the stringy foam framework championed by our previous works in~\cite{Li:2023wlo,Li:2022sgs}, supporting \textit{both} subluminal and superluminal~(i.e., CPT-violating) propagation for IceCube neutrinos. In fact, as has been mentioned as a side remark in Ref.~\cite{Li:2023wlo}~(and will be clear also from the arguments here below), one can envisage space-time foamy situations of brane defects, in which \textit{only} the deceleration of neutrinos should be expected. We therefore argue that this type of scenaria gain support from the updated result herein~\cite{Amelino-Camelia:2022pja}.

Since several suggestions for \textit{space-time foam}~\cite{Wheeler:1998stf} based on intuitive picture that the QG vacua might contain fluctuations with sizes $\sim\lp$ and timescales $t_{\textrm{P}}\sim 1/\Mp$, a number of different models that posit the appearance of Planck-scale topological fluctuations~(and, nontopological irregularities) have been proposed, ranging from microscopic~(virtual) black holes to space-time noncommutativity and defects. Resembling the example of stringy black hole~\cite{Ellis:1992eh}, a form of quantum foam may be provided by populations of the so-called \textit{D-particles}. These are stringy defects which appear in the framework of brane theories~(i.e., modern extensions to string theories with nonperturbative D(irichlet)-branes). Depending on the type of string theory considered, D-particles are either,~(i) pointlike D0-branes~\cite{Ellis:1999uh,Ellis:1999sf,Ellis:2004ay,Ellis:2008gg,Li:2021gah,Li:2022sgs,Li:2023wlo}~(type I/IIA strings), with no electric charge, in which case only charge neutral matter as well as radiation (represented as~(open) strings with ends attached on the braneworld, representing our Universe) can interact nontrivially with the foam, or,~(ii) `effectively pointlike' consisting of compactified branes, wrapped up around topologically nontrivial 3-cycles in the compact extra-dimensional manifold, permitting charge flux over the brane surfaces~(type IIB theories)~\cite{Li:2009tt}. Embedded in space-time, with four uncompactified dimensions, these nonlocal solitonic defects may be viewed as providing a sort of `environment' that induces locally a spontaneous breaking, or better, reduction of Lorentz symmetry by the quantum gravitational ground state.

In such D-brane/string scenaria, the wave propagation of a massless string mode such as a photon or graviton is affected, as originally stemming from a supercritical~(Liouville) string approach to QG~\cite{Amelino-Camelia:1996bln}, with travel times affected by amounts increasing linearly with $E$ due to an energy-dependent average shift in its phase $\sim\exp[\I\bfp\cdot{\bf x}-\I E(1+\overline{\alpha}E)t]$ with $\overline{\alpha}$ a factor encoding Lorentz violation. For the D-particle mass of order $M_{\textrm{D}}\coloneqq 1/(\ell_{s}g_{s})$, with $\ell_{s}$ the string scale, $g_{s}$ the string coupling that is assumed here to be weak, an order-of-magnitude estimate is, $\lvert\overline{\alpha}\rvert\sim\O(1/M_{\textrm{D}})$, while more information depends on the microscopic details of the interactions between the stringy matter with the `foam' of D-branes~(`\textit{D-foam}').

Initially, the preferential role of neutral gauge boson, say photon, in interactions with D-foam was established, given the gauge symmetry of electromagnetism respected by the capture/splitting process of U(1)$_{\textrm{em}}$ excitations of the Standard Model~(SM) by the D-particles. The case of SM-like neutrino could be similar, as it is, at least, not electrically charged, and an induced breaking of the electroweak SU(2)$_{\textrm{L}}$ symmetry of the SM might be expected in such situations~\cite{Li:2022sgs,Li:2023wlo}, responsible for unsuppressed foamy effects on neutrino propagation remarked below. As neutrinos are often thought to be of Majorana type one may describe instead such effects for `photinos' via supersymmetry. This may also allow to combine two Majorana~(neutrino) species into a Dirac one, if neutrinos are Dirac fermions. In any case, heuristic arguments given below do not appeal practically to supersymmetry which is not exact even at the scale of QG, and is certainly absent at lower energies, as phenomenologically required.

Hence, in one formulation of D-particle foam~\cite{Ellis:2008gg} (or a supersymmetric extension of it), when a neutrino strikes a D-particle it excites an oscillating intermediate string stretching between the defect and the D3-brane world~\cite{Ellis:2004ay} on which the neutrino propagates. The decay of the string produces the outgoing wave corresponding to the~(re)emitted neutrino, causing an infinitesimal \textit{causality-respectful} delay in the wave propagation, in a spirit reminiscent of the stretched-string formalism therein~\cite{Ellis:2008gg}, where a \textit{formal} analogy with the interaction of a wave with internal degrees of freedom in a conventional medium has been noted. Flux forces characteristic of the microscopic dynamics of D-particles in the process of capture and release are responsible for the appearance of retardation effects:
\begin{equation}
\label{eq:td}
\delta t\sim\alpha^{\prime}p^{0}/(1-\un^{2}),
\end{equation}
or, simply~\cite{Ellis:2008gg},
\begin{equation}
\label{eq:tdn}
\delta t\sim E\alpha^{\prime},
\end{equation}
as $\lvert u_{i}\rvert\ll 1$. Above, one took into account the recoil excitation of a D-particle during scattering, with $u$ its recoil velocity~(see below).\footnote{We note, for completeness, that a similar situation, with delays~(while suppressed by an extra factor $\sim\ell_{s}^{4}/{\cal V}_{4}$ depending on compactification volume ${\cal V}_{4}$) in travel times of strings, also appears to characterize the vacuum of the more complicated brane model of~\cite{Li:2009tt}, where~(while, as noted earlier~\cite{Li:2023wlo}, time advances, of same order~(\ref{eq:tdn}), may be entertained for reasons irrelevant to present-day phenomenologies) the induced \textit{subluminal} refractive index arises from direct estimate of the amplitude of scattering with the wrapped D-`particle', slowing down the propagating neutrino as here.} In the case of space-time foam with a linear density of D-defects, $n_{\textrm{D}}/\ell_{s}$, i.e., $n_{\textrm{D}}$ defects per string length, $\ell_{s}\equiv\sqrt{\alpha^{\prime}}$~(where, $\alpha^{\prime}$ is string's Regge slope), the overall delay experienced by a neutrino scattering off multiple D-branes and traversing a distance $L$~(assuming that the neutrino energy $p^{0}$ is conserved as a mean-field approximation) is of order $\Delta t\propto En_{\textrm{D}}L\sqrt{\alpha^{\prime}}$. We see that to lowest order in $p^{0}$, for neutrino species passing that `D-foamy' space-time, the refraction $n(\sim 1+L/\Delta t)$ for neutrino is independent of the neutrino chirality, and of the form
\begin{equation}
\label{eq:nri1}
0<n-1\simeq n_{\textrm{D}}^{\star}\alpha^{\prime}E\sigma_{\nu\textrm{--D}}\Bigl[\frac{g_{s}}{\ell_{s}}\Bigr]\ll 1,
\end{equation}
with unknown quantities of the theory, particularly the~(local) density of defects, taken in a \textit{uniform} D-foam model to be a constant $n_{\textrm{D}}^{\star}$, which in principle should read $n_{\textrm{D}}(z)$~\cite{Ellis:2004ay} depending on cosmic redshift $z$, and this further complicates the situation, which we do not discuss. Here, in~(\ref{eq:nri1}), $\sigma_{\nu\textrm{--D}}=\sigma_{\nu\textrm{--D}}[\ell_{s}^{-1}g_{s}]$ denotes the appropriate cross section for neutrino--D(-particle)-defect scattering.

Evidently, this feature~(\ref{eq:nri1}) points to the net effect of decelerating neutrinos~(antineutrinos), by such a `foam' of gravitational defects in space-time. Notice that physical recoil of the defects after the collisions with a neutrino is not directly relevant in obtaining the delays~(\ref{eq:tdn}), where the recoil is eliminated in case of massive D-defects, as ensured by the weak string coupling of $g_{s}\ll 1$ or just below $\O(1)$. Nonetheless, this recoil was argued~\cite{Ellis:2008gg} to resemble situations of strings in an external `electric' field, of which role is played now by $\bfu$, yielding a noncommutative geometry of space~\cite{Mavromatos:1998nz}. This indicates stringy uncertainties~\cite{Yoneya:2000bt}, related implicity to the delay; but it is also possible that the recoil, together with~(\ref{eq:tdn}) enjoys an alternative description.

Indeed, the nontrivial index of refection \textit{in vacuo} of the type~(\ref{eq:nri1}) was considered previously a result of back-reaction effects of matter excitation onto fluctuating space-time, thus of Liouville-gravity effect. In this case, recoil off such vacuum defects~\cite{Kogan:1996zv} during the scattering induces the departure from criticality~\cite{Amelino-Camelia:1996bln}, curving~(local) target space (of $D(=10)$ dimensions~\cite{Ellis:2004ay}; $A,B=1,\ldots,D-1$):
\begin{equation}
\label{eq:fmd1}
G_{AB}\sim\eta_{AB}+h_{AB},
\end{equation}
which, asymptotically far away from the~(recoiling) defects, resembles a perturbed gravitational field of Finsler~(momentum-dependent) type
\begin{equation}
\label{eq:fmd2}
G_{0i}=h_{0i}\simeq u_{i\Vert},\quad i=1,2,3.
\end{equation}
Here, $u_{i\Vert}=\ell_{s}g_{s}\Delta p_{i}$ denotes the recoil~(parallel) velocity of the D-particle given a momentum transfer, $\Delta p_{i}\equiv p_{i}\zeta_{\Vert}$~($\zeta_{\Vert}<1$) from the incident momentum $\bfp$ of the particle. Technically, utilizing an appropriate supersymmetrization~\cite{Ellis:1999sf} of the Born--Infeld(-type) action initially used to treat the recoil of bulk brane defect after scattering off bosonic matter, say, photon~\cite{Amelino-Camelia:1996bln,Mavromatos:1998nz,Ellis:1999uh}, allows to establish a same effect~(\ref{eq:fmd2}) for photino field~(a Majorana spinor). One then has the modified Dirac equation:\footnote{According to the analysis of the low-energy effective field theory~(EFT) in the D-particle `gravity media'~\cite{Ellis:1999sf}, we could write the kinetic part of the pertinent Dirac action in the distorted background~(\ref{eq:fmd2}), in the more suggestive form~(see,~Appendix~\ref{appb})
\begin{align*}
-S&=\int\d^{4}x\sqrt{-G}\,\overline{\Psi}( G_{\alpha\beta}\Gamma^{\alpha}\D^{\beta}+m_{\nu})\Psi\tag{$\star$}\label{eq:mda}\\
&=\int\d^{4}x(1+u_{i}u_{j}\delta^{ij})^{\frac{1}{2}}\overline{\Psi}(\slashed{\partial}- u_{i}\gamma^{0}\partial^{i}+\O(u^{2})+m_{\nu})\Psi.
\end{align*}
This involves the vierbein $\hat{e}^{m}$ explicitly, with the~(bare) mass $m_{\nu}$. The time delay of the linear form~(\ref{eq:tdn}) may be interpreted in this case as a result of the induced average dispersion~(\ref{eq:andr}) but \textit{not necessarily}, since the formalism of~\cite{Ellis:2008gg}, giving rise to delays~(\ref{eq:tdn}), which are essentially stringy~(uncertainty) effects, basically, does \textit{not} admit a local EFT~(\ref{eq:mda}) treatment.}
\begin{equation}
\label{eq:mde}
(\gamma^{m}\partial_{m}-\gamma^{0}\bfu_{\Vert}\cdot\nabla+ m_{\nu})\Psi=0,
\end{equation}
where the mass term $m_{\nu}$ can be omitted in the case of relativistic neutrinos, and Dirac $\gamma^{m}$ matrices are associated to that $\Gamma^{\alpha}$ in general relativity via $\gamma^{m}=e^{m}{\!}_{\alpha}\Gamma^{\alpha}$, with $\{\Gamma^{\alpha},\Gamma^{\beta}\}\coloneqq\Gamma^{\alpha}\Gamma^{\beta}+\Gamma^{\beta}\Gamma^{\alpha}=2G^{\alpha\beta}\mathds{1}$, $\{\gamma^{m},\gamma^{n}\}=2\eta^{mn}\mathds{1}$, as usual, and $\hat{e}^{m}(u)$, a vierbein (tetrad) taking the form, e.g., of
\begin{equation}
\label{eq:vc}
e^{\hat{0}}{}_{0}=1,\quad e^{\hat{0}}{}_{i}=u_{i},\quad e^{\hat{i}}{}_{\beta}=\eta^{\hat{i}}{}_{\beta}.
\end{equation}
Here $\hat{i}$ denotes the index of flat Minkowski space $\hat{\eta}$. It is easy to derive a quadric-order differential equation from equation~(\ref{eq:mde}), which describes formally the propagation for the neutrinos and their charge conjugates~(i.e., antineutrinos; or, Majorana ones with opposite chiralities), $\Psi=(\nu_{e},\nu_{\mu},\nu_{\tau},\nu_{e}^{c},\nu_{\mu}^{c},\nu_{\tau}^{c})$:
\begin{equation}
\label{eq:sde}
[\partial_{t}^{2}-\nabla^{2}+2(\bfu_{\Vert}\cdot\nabla)\partial_{t}+m_{\nu}^{2}]\Psi\simeq 0,
\end{equation}
With the appropriate use of the plane wave \textit{Ansatz}: $\Psi(x)\sim\e^{\I p\cdot x}$ the dispersion relation reads
\begin{equation}
\label{eq:ndr}
-E^{2}+2E\bfp\cdot\bfu_{\Vert}+\bfp^{2}\sim 0,
\end{equation}
analogous to the one possessed by an energetic photon from similar analyses of the modified Maxwell's equation~\cite{Amelino-Camelia:1996bln} and/or on-shell condition~\cite{Ellis:2004ay,Li:2021gah}~(i.e., $p^{\alpha}G_{\alpha\beta}p^{\beta}=-m^{2}$) or, of the null geodesic~\cite{Ellis:1999uh}~(as is appropriate for~(almost) massless neutrino) in such a disturbed isotropic space-time~(\ref{eq:fmd1}).

Following up from above alternative formulation, we may consider then, a collection of D-particles encountered by the neutrino, leading to an \textit{anisotropic} foam situation, by which we mean an average background recoil-velocity field $\dla u_{i}\dra\neq 0$, provided that the momentum transfer variable $\zeta_{\Vert}$ in~(\ref{eq:fmd2}), is nonvanishing, upon taking the average $\dla\cdots\dra$ over ensembles of bulk defects $0\neq\dla\zeta_{\Vert}\dra\eqqcolon\zeta_{\textrm{D}}$, with $\zeta_{\textrm{D}}\lesssim\O(1)$ a dimensionless numerical parameter, proportional to the D-defect density, that is $\zeta_{\textrm{D}}\equiv {\cal C}n_{\textrm{D}}^{\star}$, with ${\cal C}$ a fudge factor. We obtain, then, upon averaging~(\ref{eq:ndr}),\footnote{Bear in mind that, henceforth, for simplicity, we omit the notation $\dla\cdots\dra$'s, i.e., expectation values in the foam vacuum, since all the quantities such as $p_{i}$, $E$ and the velocity $v$ below denote such averages.} modified dispersions valid for either Dirac or Majorana neutrinos~(at high energies: $m_{\nu}/E\rightarrow 0$):
\begin{equation}
\label{eq:andr}
E^{2}\simeq\bfp^{2}\Bigl(1-2{\cal C}n_{\textrm{D}}^{\star}\frac{g_{s}E}{M_{s}}+\ldots\Bigr),
\end{equation}
where $M_{s}\equiv\ell_{s}^{-1}$, and the minus sign in~(\ref{eq:andr}) reflects the fact that Eq.~(\ref{eq:mde}) entails \textit{no} helicity dependence. The analysis that leads to that relation goes in parallel with that done in the bosonic~(say, a photonic) case~\cite{Ellis:2004ay,Li:2021gah}. Hence the low-energy target-space recoil dynamics governed by the fermionic analogy of a Born--Infeld treatment~\cite{Mavromatos:1998nz} keeps the propagation \textit{subluminal}, with a refractive index:
\begin{equation}
\label{eq:nri2}
n=1+\O(\zeta_{\textrm{D}(\nu)}\pn/M_{\textrm{D}})>1.
\end{equation}
Here, the suffix $(\nu)$ denotes explicitly a parameter $\zeta$ for an ordinary neutrino~(antineutrino), because the effects of such a QG environment differ for different particle types~\cite{Li:2022ugz}~(considering a plausible violation of the equivalence principle~\cite{Ellis:2003sd}; see also, Ref.~\cite{Li:2023wlo}), as stated before.~(Of course \textit{isotropic} foam scenaria with $\dla u_{i\Vert}\dra=0$ may also be envisaged, affecting the dispersion by a subleading amount, instead of~(\ref{eq:andr}), but one should still expect uncertainty-caused time delays~(\ref{eq:tdn}); see below.)

Whether Eq.~(\ref{eq:nri1}), or~(\ref{eq:nri2}) illustrate that the phase velocity for a neutrino wave, $v_{\textrm{ph}}\coloneqq 1/n$, is \textit{subluminal}, as is in principle the group velocity, $v$. Indeed, to obtain the latter, one can adopt the derivative of the index of refraction with respect to $E$,
\begin{equation}
\label{eq:dri1}
\frac{1}{n}=\Bigl(1+\pn\frac{v_{\textrm{ph}}}{n}\frac{\d n(E)}{\d E}\Bigr)v,
\end{equation}
which, in the case of a novel quantum-gravitational `\textit{medium}', with an induced index of refraction given by $n-1=\chi_{\textrm{QG}}E$~(with $\chi_{\textrm{QG}}$, a constant, as we have exactly here), leads to
\begin{eqnarray}
\label{eq:dri2}
& &\frac{1}{v}=n+\chi_{\textrm{QG}}E=1+2\chi_{\textrm{QG}}E\nonumber\\
& &\Rightarrow v\simeq 1-2\chi_{\textrm{QG}}E<1,
\end{eqnarray}
if $2\chi_{\textrm{QG}}E\ll 1$ and $\chi_{\textrm{QG}}>0$. Hence, the \textit{subluminal} effective~(group) velocity $v$ for the high-energy neutrinos/antineutrinos~(with negligible mass, $m_{\nu}\simeq 0$) traversing a dispersive QG medium of the type~(\ref{eq:nri2}) yields here,
\begin{equation}
\label{eq:ngv}
1-v^{\textrm{D-foam}}\approx\varsigma_{(\nu)}\ell_{s}E,
\end{equation}
where $\varsigma$ is some fudge factor, entailing information on the momentum $p_{i}$ transfer during the scattering, as encoded in $\dla u_{i}\dra\sim\zeta_{\textrm{D}}\eqqcolon\varsigma/(2g_{s})$, and it seems to have the form $\varsigma\sim n_{\textrm{D}}^{\star}\ell_{s}\sigma_{\nu\textrm{--D}}$ in the space-time-foam formulation of~\cite{Ellis:2008gg} which gives rise to the refractive index effect characterized by~(\ref{eq:nri1}). 

Notice however that the derivation of~(\ref{eq:nri2}--\ref{eq:dri2}) relies on the usual expression, $v=\partial E/\partial\pn$, for which the validity in the context of QG is still in dispute. On assuming that it holds the resulting retardation of neutrinos with observed energies $E$ due to linear modification of LV~(\ref{eq:ngv}) is of the form
\begin{equation}
\label{eq:ntd}
\Delta t^{\textrm{D-foam}}\simeq \frac{E}{M_{\textrm{sQG}}}L,
\end{equation}
as they travel over a distance of $L$, taking account of the cosmic expansion~(below, $H(z)$ and $H_{0}$ indicate the Hubble parameter and the Hubble constant, respectively),
\begin{equation}
\label{eq:cd}
L(z)=H_{0}^{-1}\int_{0}^{z}\frac{(1+{\cal Z})\d{\cal Z}}{H({\cal Z})}.
\end{equation}
We may write here the gravity foam mass characteristic of the D-brane approach to stringy QG, in~(\ref{eq:ntd}) as $M_{\textrm{sQG}}\coloneqq1/(\sqrt{\alpha^{\prime}}\varsigma)$, which, in an `anisotropic' D-foam Universe where $\dla u_{i}\dra\neq 0$, takes a finite value, to be determined experimentally.\footnote{Certainly this QG scale diverges in the limit of vanishing density of D-particles and/or~(in our previous foam model thereof~\cite{Li:2023wlo,Li:2022sgs}) zero fluctuations of $u_{i}$~(evaluated over a stochastic ensemble of D-defects), as expected, restoring Lorentz symmetry.} Even if $\dla u_{i}\dra=0$ (`isotropic recoil' foam), nonetheless, such a LV delay~(\ref{eq:ntd}) arises, as a result of~(\ref{eq:tdn}), from accumulating uncertainties of the string stretched between the D-brane defects of the foam with the braneworld over a neutrino travel 
distance $L$, with $M_{\textrm{sQG}}\propto M_{s}/n_{\textrm{D}}^{\star}$. This implies the disentanglement of delays~(\ref{eq:ntd}) linear in neutrino energy from the modified dispersion relations in that case, 
\begin{equation}
\label{eq:nndr}
E^{2}-\bfp^{2}\sim\O\Bigl(+\,\bfp^{2}\frac{E^{2}}{M_{\textrm{D}}^{2}}\Bigr),
\end{equation}
whose corrections scale quadratically with energy $E$ due to higher-order~(stochastic) processes~\cite{Mavromatos:2005bu} in the foam vacuum~(i.e., $\dla u_{i}u^{i}\dra\neq 0$, as one expects~\cite{Li:2023wlo}); as such we may consider it as providing an example pointing to a possible failure of the velocity law~(\ref{eq:dri1}) in the scenaria of~\cite{Ellis:2008gg,Li:2009tt}.

Therefore, what we would like to stress is that the central feature of either of the above descriptions is always \textit{the appearance of a linear `delay'~(\ref{eq:ntd}), irrespective of the chirality status, in propagation times of neutrinos}. So, in any case, energetic cosmic neutrino~(antineutrino) would decelerate in foam types assumed in this article, thereby compatible with the phenomenological indication~(from recent Ref.~\cite{Amelino-Camelia:2022pja}) that, the recalibrated data of IceCube seemingly favor more the GRB neutrinos that might have been slowed down by LV, as mentioned at the beginning. This in turn lends support to the formalism we prefer here for string theory foam.

An important comment arises here, to avoid some confusion, regarding testing predictions of the models: the clearest test of such models is the one where one simply establishes delays in travel times of neutrinos and checks their dependence on particle energies, as has been done by~\cite{Amelino-Camelia:2022pja}~(or in previous~\cite{Amelino-Camelia:2016fuh,Amelino-Camelia:2016ohi,Huang:2018ham,Huang:2019etr,Huang:2022xto}). The precise magnitude of the effects is controlled by the mass $M_{\textrm{sQG}}$. One might expect $M_{\textrm{sQG}}=\O(\Mp)$ whilst it depends on details of the foam, such as the D-particle density $n_{\textrm{D}}^{\star}$ and the string scale $\ell_{s}$~(which $\neq\lp$, in general). Such details cannot be estimated via theoretical considerations at present, due to our imperfect knowledge of microscopic theory at a very fundamental level. Consequently, $n_{\textrm{D}}^{\star}$ is an arbitrary free parameter in this context, so that, $\varsigma$~(or, $M_{\textrm{sQG}}$) is actually free to vary, and needs to be constrained phenomenologically.

Based on the above discussion we are now able to make an estimate of the possible values that the QG parameters could take in order to match the observations updated by~\cite{Amelino-Camelia:2022pja}, i.e., to produce the intriguing feature found in the revised IceCube data~\cite{IceCube:2020wum}: translating the result from the best-fit for $\Elv$~(for $s=+1$) to $M_{\textrm{sQG}}$, we have
\begin{equation}
\label{eq:qgv}
\frac{\varsigma_{(\nu)}}{g_{s}}\sim\O(10^{-18})\frac{M_{\textrm{D}}}{\textrm{GeV}},
\end{equation}
or simply $M_{s}/n_{\textrm{D}}^{\star}\sim 10^{18}$~GeV. This implies natural values for both parameters. If, as traditional, $M_{s}$ is of or below, the order of the~(4-dimensional) Planck mass, say, $\leq 10^{16}$~GeV, we infer an upper bound for the dimensionless factor, $\varsigma$~(with $g_{s}<1$) in~(\ref{eq:ntd}), for illustrative purposes:
\begin{equation}
\label{eq:fpv}
\varsigma_{(\nu)}\lesssim 10^{-2},
\end{equation}
which is precisely consistent with our perception for D-foam models, $\varsigma\leq\O(1)$ for $\varsigma$, characteristic of microscopic interactions of the neutrino with multiple defects.\footnote{Recall that the recoil effect envisaged above is essentially geometrical and kinematical and as such the D-foam is \textit{flavor blind}~\cite{Ellis:1999sf,Li:2023wlo}, i.e., $\varsigma$~(or, $M_{\textrm{sQG}}$) protects the flavor symmetry, obviating any impact on neutrino oscillation physics~(despite of the presence of stochastic fluctuations~\cite{Ellis:1999uh} in the velocities of mono-energetic neutrinos, with a \textit{diffusive} character which is but still beyond observability), in contrast to that asserted by~\cite{Lambiase:2001ib} based upon the premise that flavor eigenstates couple in different ways to the metric of~(\ref{eq:fmd2}). This is \textit{not} the case~\cite{Li:2023wlo} in the string model(s) we are working with.} Nevertheless, it must be noted again that, according to modern version of strings $M_{s}$ is in general arbitrary, and it can be as low as a few TeV~(in several schemes~\cite{Antoniadis:1999rm} of low-scale strings), depending on compactification procedure, which unfortunately entails all sorts of complications coming from supersymmetry breaking. In such cases, very weak string couplings $g_{s}$, or $\varsigma\ll 1$, would be required, such that a D-foam-vacuum refractive effect, of a strength appropriate to generate delays inferred by~\cite{Amelino-Camelia:2022pja}, slowed those neutrinos down. Even there, the order of relevant quantity, $\zeta_{\textrm{D}} g_{s}$~(or, simply $\varsigma$), depending on the effective density of D-defects $n_{\textrm{D}}^{\star}$ which would be as low as $\sim 10^{-14}$, is within reasonable ranges, considering~\cite{Ellis:2004ay,Ellis:2008gg,Li:2021gah} that this latter quantity is another free parameter of such brane models, as we have already emphasized. In such cases of low $M_{s}\geq\O(10)$~TeV, one may easily envisage a much more dilute population of D-defects in the bulk, so that their densities on the branes fall naturally below such values. But, of course, we have to keep an open mind about values out of the result~(\ref{eq:qgv}), especially in view of more refined analysis in the future with many more data, as compared to the present one~\cite{Amelino-Camelia:2022pja}.

In `foamy' situations considered here, where only subluminal propagation is present, any LV bounds, relevant for~(putative) superluminal velocities~(such as the one established by~\cite{Wang:2020tej} from the relatively robust evidence~\cite{IceCube:2018dnn} that a neutrino with energy of at least $\sim 183$~TeV was observed from TXS~0506+056, and not depleted by anomalous processes like, $e^{+}e^{-}$ pair production \textit{in vacuo}) are naturally evaded, but in several different ways:
\begin{itemize}
\item In the anisotropic D-foam~(\ref{eq:andr}),~(\ref{eq:nri2}), the analysis of the pertinent reaction~(e.g., $\nu\rightarrow\nu e^{+}e^{-}$), based on a putative superluminal dispersion relation for the neutrino is, obviously, inapplicable, as a result of the subluminal nature of the average recoil phenomenon.
\item For the stretched-string and/or isotropic recoil models, if $\O(\textrm{eV}^{2})\sim m_{\nu}^{2}\ll\lvert\bfp E\rvert\ll (M_{s}/g_{s})^{2}$, Eq.~(\ref{eq:nndr}) looks roughly like a relativistic dispersion relation $E\sim\pn$, suggesting the \textit{intactness} of the kinematics of neutrino-related channels; but there still exist \textit{causal} delays~(\ref{eq:tdn}), which are associated~\cite{Ellis:2008gg} with purely stringy~(and hence, nonlocal) effects, not captured by the neutrino in-vacuo dispersions. Keeping terms of higher-order in such induced dispersions~(\ref{eq:nndr}) does allow one to bound, e.g., anomalous~(`superluminal', by which we mean a `+' sign thereof) corrections quadratically suppressed by the~(new) `QG dispersion scale', due to the stochasticity condition~($\dla u_{i\Vert}u_{j\Vert}\dra\sim\delta_{ij}\dla\zeta_{\Vert}^{2}\dra\neq 0$):
\begin{equation}
\tilde{M}\sim\frac{M_{s}}{g_{s}\sqrt{\dla\zeta^{2}\dra}}\propto M_{\textrm{D}}/\sqrt{n_{\textrm{D}}^{\star}},
\end{equation}
a case for which constraints coming from high-energy cosmic neutrinos are currently weak: as examples, the results of~\cite{Wang:2020tej} may be thought of implying a bound of order $\tilde{M}\gtrsim 10^{13}$~GeV; but of course, the situation may change with future experiments.
\end{itemize}

It should be emphasized that arguments concerning string models adopted here to confront the possible \textit{enhancement} of observational hint~\cite{Amelino-Camelia:2022pja} for delayed GRB neutrinos, as provided above, are not as that trivial as one might first imagine. In fact as far as the latest revision of IceCube neutrino directions and its consequences~\cite{Amelino-Camelia:2022pja} for previous LV-motivated analyses~\cite{Amelino-Camelia:2016fuh,Amelino-Camelia:2016ohi,Huang:2018ham,Huang:2019etr,Huang:2022xto} are concerned, such models could be favored over the CPT-breaking one~\cite{Li:2023wlo,Li:2022sgs} that was recently formulated in the D-foam framework. This is the central point we would like to clarify here, and regretfully, has not been discussed in~\cite{Li:2023wlo,Li:2022sgs}, where emphasis was lain upon the attempts to explain previous suggestion of neutrino speed variation of possibly CPT-violating form, $v=1\mp E/\Elv$~(inferred from uncorrected neutrino data).

Nevertheless we note that while only part of what has been suggested previously in Refs.~\cite{Amelino-Camelia:2016fuh,Amelino-Camelia:2016ohi,Huang:2018ham,Zhang:2018otj,Huang:2019etr,Huang:2022xto} might be physical, with the rest~(the `superluminal' part), being likely just accidental results, it renders a comparable $\Elv$ for subluminal propagation with linear dependence on neutrino energy as in~\cite{Amelino-Camelia:2022pja}; and again one can arrive at similar estimate~(\ref{eq:qgv}) as above, albeit the propagation properties phenomenologically speculated are different.

In closing we remark on the claim~\cite{Amelino-Camelia:2022pja}, that~(with the recalibrated IceCube data) there seems now no, or better, little, support for `advance' events which correspond to neutrinos~(antineutrinos) undergoing acceleration by the effect of QG: while the reported probability of finding coincidentally at least 3 such events from 60--500~TeV shower neutrinos is as high as 81\%~\cite{Amelino-Camelia:2022pja}~(implying that they are probably to be just pure background) one cannot exclude that this might have to be reassessed with more IceCube neutrino data being collected. Also, it seems to us that support from the presently available data for the coexistence of both time `delay' and `advance' events has not evaporated completely yet, not even models of space-time foam consistent with such indications say, the one considered recently in our Refs.~\cite{Li:2023wlo,Li:2022sgs}. The CPT-violating vacuum refraction for neutrinos from that model would seem likely to raise its feasibility again, if the validity of the superluminal part of the data\footnote{We emphasize that the argument that superluminal neutrinos should be strongly diminished by, e.g., $\nu\rightarrow\nu e^{+}e^{-}$ (or, $\overline{\nu}\rightarrow\overline{\nu} e^{+}e^{-}$), as cited by~\cite{Amelino-Camelia:2022pja} as a consistent support for their findings of only subluminal GRB neutrinos, is \textit{not} applicable in the context of the D-brane model of~\cite{Li:2022sgs}, because superluminal~(anti)neutrinos could propagate~(practically-)stably in the QG foam of that type~\cite{Li:2023wlo}. Constraints from $\overline{\nu}\rightarrow\overline{\nu} e^{+}e^{-}$ may be avoided, again, but by this means.} were to be strengthened.

To sum up, given the importance of uncovering stronger evidence~\cite{Amelino-Camelia:2022pja} for neutrinos slowed down by LV~(while weakening the signature of those events of the `advance' type formerly suggested~\cite{Amelino-Camelia:2016fuh,Amelino-Camelia:2016ohi,Huang:2018ham}) due to the recent correction~\cite{IceCube:2020wum} of IceCube data, it is useful to follow up on this updated indication, with which certain~(stringy) QG model may be put to test. Following early treatment~\cite{Ellis:1999sf} of a fermionic system in space-times distorted by a recoiling D-brane, we extend this formalism, regarding the foamy aspects of such models, to situations involving a large population of D-particles~(leading to average effects~(\ref{eq:nri2})). The refractive index found for neutrino is necessarily subluminal and consistent with the one~(\ref{eq:nri1}) from the microphysical model~\cite{Ellis:2008gg}, boiling all down to delays in its propagation; this is to be contrasted with the class of models of~\cite{Li:2023wlo,Li:2022sgs}, where the appearance of both delayed and, advanced~(anti)neutrinos may be expected, and can then explain the previous~(but now seemingly less favored) speculation of CPT violation in cosmic neutrinos using uncorrected event directions. This is an important feature of either of the two formulations preferred here for the D-foam. We therefore argue that this sort of model could be in agreement with the phenomenological suggestion of neutrino slowing-down~\cite{Amelino-Camelia:2022pja}, at scales of order~(at least not lower than) $10^{17}$~GeV. An extended study on the recalibrated IceCube data, in addition to the reanalyzed $\gtrsim 60$~TeV neutrinos, will further test~(or even push the limits of) our insight.

\section*{Acknowledgments}

This work was supported by the National Natural Science Foundation of China (Grants No.~12075003 and No.~12335006). C.L. was supported in part by a Boya Fellowship provided by Peking University, and by the Postdoctoral Fellowship Program of CPSF under Grant No.~GZB20230032.

\appendix

\section{Metric conventions}
\label{appa}

We adopt here the convention often taken in relativity, gravity, and~(super)string communities, that is, the `east coast', or, `mostly plus' metric convention $\eta_{AB}=\textrm{diag}[-1,+1,\ldots,+1]$, in which case~(the chiral basis for) the Dirac matrices:
\begin{subequations}
\begin{equation*}
\gamma^{0}=\biggl(
\begin{array}{cc}
0 & -\I\mathds{1}_{2\times 2}\\
-\I\mathds{1}_{2\times 2} & 0
\end{array}\biggr),\gamma_{k}=\biggl(
\begin{array}{cc}
0 & -\I\hat{\sigma}_{k}\\
\I\hat{\sigma}_{k} & 0
\end{array}\biggr),
\end{equation*}
\begin{equation}
\mathds{1}_{2\times 2}\equiv\mathds{1}=\biggr(
\begin{array}{cc}
1 & 0\\
0 & 1
\end{array}\biggr),\tag{A.1}
\end{equation}
\end{subequations}
where the two-by-two matrices, $\hat{\sigma}_{k}$, with $k=1,2,3$, denote the usual Pauli~(spin) matrices. The conventional Dirac equation, whose indices are contracted with the~(4-dimensional) flat Minkowski space-time background metric $\hat{\eta}$,~($\eta_{\alpha\beta}$), as is understood, takes accordingly the form
\begin{equation}
(\gamma\cdot\partial+m)\psi=0,
\end{equation}
\noindent which arises now from the fermionic kinetic term in the Lagrangian~(below, $\slashed{\partial}\equiv\gamma\cdot\partial$):
\begin{equation}
-\overline{\psi}(\slashed{\partial}+m)\psi.
\end{equation}
These differ slightly from the form they would take if we adopted a Lorentzian signature that is mostly minus, but they have exactly the same content. The reader should be aware of that notation.

\section{Derivation of~(\ref{eq:ndr}) given the recoil of a \textit{single} D-particle and, remarks}
\label{appb}

Coupling gravity to spin-$\frac{1}{2}$ matter requires a generalization of the flat-space Dirac action; the slight subtlety is that the spinor field is defined relative to the locally Minkowskian~(thus, local Lorentz) frame specified by the vierbein $\hat{e}^{m}$. Hence,
\begin{equation}
S=-\int\d^{4}x\sqrt{-G}\,\overline{\psi}(\gamma^{m}\eta_{mn}e^{n\alpha}\D_{\alpha}+m)\psi,
\end{equation}
where $\d x^{4}\sqrt{\lvert G\rvert}$ indicates the volume element~($G\coloneqq\det(G_{\alpha\beta})$; the covariant derivative has to be defined in terms of the connection, $\hat{\omega}$,~($\omega_{mn\alpha}=e_{m}{\!}^{\beta}e_{n\beta;\alpha}$): $\D_{\alpha}\psi=(\partial_{\alpha}+\frac{1}{4}\omega_{mn\alpha}\gamma^{[mn]})\psi$, with $e_{n\nu;\alpha}\equiv\partial_{\alpha}e_{n\beta}-{}^{c}\Gamma_{\beta\alpha}^{\gamma}e_{n\gamma}$~(here, ${}^{c}\Gamma$ is the Christofell symbol relative to $G_{\alpha\beta}$) and $\gamma^{[mn]}\coloneqq\frac{1}{2}[\gamma^{m},\gamma^{n}]$. From this position follows that~(c.f.~Eq.~(\ref{eq:mda})),
\begin{equation}
-\tilde{S}=\int\d^{4}x\,e\Bigl\{\frac{1}{2}(\overline{\psi}\overset{\text{\tiny$\leftrightarrow$}}{\slashed{\D}}\psi)+m\overline{\psi}\psi\Bigr\},
\end{equation}
which is equivalent to $S$~(that is $(S/2+\textrm{h.c.})\mapsto \tilde{S}$); $G=\det(ee\eta)=-[\det(e^{m}{\!}_{\alpha})]^{2}\eqqcolon -e^{2}$, $\slashed{\D}\coloneqq\Gamma\cdot\D=\Gamma^{\beta}G_{\beta\alpha}\D_{\alpha}$, and $\overset{\text{\tiny$\leftrightarrow$}}{\D}_{\alpha}=\overset{\text{\tiny$\rightarrow$}}{\D}_{\alpha}-\overset{\text{\tiny$\leftarrow$}}{\D}_{\alpha}$.

Consider the foam-like Finsler metric~(\ref{eq:fmd2}) taken to be small perturbation about $\eta_{\alpha\beta}$, i.e., $h_{0i}=u_{i}\ll 1$, then, upon using the above formalism, equations of motion read Eq.~(\ref{eq:mde}). This was first obtained by~\cite{Ellis:1999sf} in the context of D-brane recoil induced by a \textit{single} scattering event with the propagating fermion~(say, neutrino) field in asymptotically flat space-time~(\ref{eq:fmd1}). We apply the usual trick by acting twice with foam-modified `Dirac operator'~($\gamma\cdot\partial+m=\gamma^{0}\bfu\cdot\nabla$) to get the dispersion. Writing this out yields
\begin{align}
0&=\{\gamma^{n}\gamma^{m}\partial_{n}\partial_{m}-\{\gamma^{n},\gamma^{0}\}\bfu_{\Vert}\cdot\nabla\partial_{n}+(\bfu_{\Vert}\cdot\nabla)^{2}\}\Psi\nonumber\\
&\simeq\{\partial_{m}\partial^{m}-\gamma^{n}\gamma^{0}(\bfu_{\Vert}\cdot\nabla)\partial_{n}\}\Psi,
\end{align}
by noticing that $\gamma^{n}\partial_{n}=\gamma^{0}u_{i}\partial^{i}$~(ignoring now small mass $m_{\nu}$, as is appropriate for our discussion about $\O(\textrm{TeV}-\textrm{PeV})$ GRB-neutrino in the text), and that $\lvert u_{i}\rvert\ll 1$ validates this perturbative expansion. We have~(\ref{eq:ndr}) from above~(c.f.~Eq.~(\ref{eq:sde})):
\begin{equation}
(\bfu_{\Vert}\cdot\nabla)\partial_{t}\Psi\simeq\frac{1}{2}(\square^{2}+\O(u_{\Vert}^{2}))\Psi,
\end{equation}
where, again, we used ${\bm\gamma}\cdot\nabla=-\gamma^{0}(\partial_{t}-\bfu\cdot\nabla)$ and $\square^{2}\equiv\partial^{m}\partial_{m}=\nabla^{2}-\partial_{t}^{2}$ is the d 'Alembert operator. The quantification of~(\ref{eq:andr}) which expresses the \textit{average} effect due to scattering off multiple defects with nonvanishing $\dla u_{i\Vert}\dra$~(according to the prescriptions of Refs.~\cite{Ellis:2004ay,Li:2021gah} in the photonic case) is straightforward then, and omitted here.

We end with a technical comment about the form invariance of~(\ref{eq:mde})~(or of~(\ref{eq:andr})), dictated by the~(string theoretical) ideas underlying such microscopic foam models, through demanding that $u_{i}$ is embedded in a vierbein $\hat{e}^{m}$ with components~(\ref{eq:vc}) in each reference frame. In effect, there is a \textit{reduced symmetry} which\footnote{This is the reason why in this context, \textit{Lorentz reduction}, rather than hard violation, is expected, as stated in the text.} characterizes these scenaria, i.e., a subgroup of the SL$(2,\,\IC)$ symmetry~\cite{Ellis:1999sf}, leaving invariant the norm of the recoil 3-velocity $\lvert\bfu_{\Vert}\rvert$. Lorentz covariance is of course preserved when one entertains stochastically fluctuating foam situations with a zero mean of $u_{i}$, as is conjectured to feature also the stringy model\footnote{It has been speculated~\cite{Li:2022sgs} that the appearance of a novel type reduced-Lorentz symmetry, which preserves covariance, might serve as the reasoning behind the `inhibition' of superluminal $\overline{\nu}$-decays in that model.} of~\cite{Li:2022sgs,Li:2023wlo}. In the one development of the approach, as adopted here, we end up with an induced vacuum refraction for neutrinos~(\ref{eq:nri2}) as well as the resulting delay, of the form~(\ref{eq:ntd}).

In comparison of the refractive indices~(\ref{eq:nri1}), we reiterate that such expressions hold even if $\dla u_{i}\dra=0$, as can be seen from~(\ref{eq:td}); this is because they are \textit{not} directly related to the deformed dispersion relation of~(\ref{eq:andr}) which no longer holds in case of zero average recoil and is replaced by~(\ref{eq:nndr}) of $\O(M_{s}^{2})$ suppression, and because $\delta t$ of~(\ref{eq:tdn})~(which is linear in the energy) in this case is a purely stringy effect~(derived within the language of critical string theory)~\cite{Ellis:2008gg}, being associated with the~(stringy) time-space uncertainties $\delta X\delta t\geq\alpha^{\prime}$~\cite{Yoneya:2000bt} by
\begin{equation}
\delta P\delta X\geq 1+\O(\alpha^{\prime}).
\end{equation}
\noindent But nonetheless, in either case there are \textit{delays}~(\ref{eq:ntd}) in the travel times of more energetic neutrinos compared to less energetic ones. As argued in our work, these might be phenomenologically desirable, as far as the indications from~\cite{Amelino-Camelia:2022pja}~(and potentially~\cite{Amelino-Camelia:2016fuh,Amelino-Camelia:2016ohi,Huang:2018ham,Huang:2019etr,Huang:2022xto}) are concerned.

\end{document}